\documentclass[acmsmall]{acmart}

\AtBeginDocument{%
  }

\setcopyright{acmlicensed}
\copyrightyear{2024}
\acmYear{2024}
\acmDOI{XXXXXXX.XXXXXXX}

\acmConference[CSCW '24]{The 27th ACM Conference on Computer-Supported Cooperative Work and Social Computing}{November 09--13,
  2024}{San Jose, Costa Rica}
\acmISBN{978-1-4503-XXXX-X/24/09}

\begin{document}

\title{Contesting Artificial Moral Agents}

\author{Aisha Aijaz}
\email{aishaa@iiitd.ac.in}
\orcid{0000-0002-8137-106X}
\affiliation{%
  \institution{IIIT-Delhi}
  \city{New Delhi}
  \state{Delhi}
  \country{India}
}

\renewcommand{\shortauthors}{Aisha Aijaz}

\begin{abstract}
There has been much discourse on the ethics of AI, to the extent that there are now systems that possess inherent moral reasoning. Such machines are now formally known as Artificial Moral Agents or AMAs. However, there is a requirement for a dedicated framework that can contest the morality of these systems. This paper proposes a 5E framework for contesting AMAs based on five grounds: ethical, epistemological, explainable, empirical, and evaluative. It further includes the spheres of ethical influences at individual, local, societal, and global levels. Lastly, the framework contributes a provisional timeline that indicates where developers of AMA technologies may anticipate contestation, or may self-contest in order to adhere to value-aligned development of truly moral AI systems.
\end{abstract}




\maketitle

\section{Introduction}
\label{1}
AI technology is advancing at a rapid rate, and so is its ubiquity. However, the methods of regulating AI are left far behind, to the extent that the regulations that do exist may not be enforceable in the face of true harm \cite{smuha2021race}. This a preliminary case for why AI must be contested and upheld to predefined standards \cite{kaminski2021right}. A niche in the vast space of AI focuses on intrinsically moral AI, or \textit{artificial moral agents} (AMAs). These systems are not to be confused with AI systems that follow ethical or value-aligned engineering, although this is an important requirement for all AI systems \cite{spiekermann2023value}. AMAs are AI systems that claim to make moral decisions, or at least ethically-informed ones. Whether or not AMAs are considered moral agents is debatable \cite{formosa2021making, brozek2019can}, given they are not considered legal persons \cite{dremliuga2019criteria}. Their lack of sentience and accountability of AMAs render them void of legal personhood \cite{dremliuga2019criteria}, thus requiring a separate, more dedicated consideration to contest their apparent morality \cite{llorca2024moral}. The question then arises, \textit{how may one contest an AMA?} Perhaps we may use the law, as it is a representation of the moral consensus of a particular geopolitical area \cite{gonthier2003law}.  It considers local context, thus enforcing a sort of deontic reasoning to uphold ethical principles within its applicable boundaries. However, due to a significant lack of laws and regulation for AI systems \cite{white2022ignorance}, there is a need for a framework to facilitate their contestation. 

This paper proposes the 5E framework to fill this gap, which allows a third party to contest the morality of artificial moral agents based on a derived sphere of influence model. The framework proposes five grounds for contestation, which consist of various aspects that an AMA must deliver on, in order to be considered acceptably moral.

\section{Background}
\label{2}
Artificial moral agents are a subclass of AI systems that are able to make moral decisions without human intervention or have moral reasoning capabilities \cite{van2002deontic, formosa2021making}. The IEEE 7000 standard \cite{ieee} aims to address ethical concerns during system design, akin to the work of Alfrink et al., who have also provided a framework for contesting AI systems by design \cite{alfrink2023contestable}. The framework proposed in this paper aims to tailor these contributions to AMA systems.

When building a framework for the moral contestability of AMAs, we must exclusively consider those machines that claim to make autonomous moral decisions. According to Moor's classification for ethical agents, only \textit{explicit ethical agents} \cite{moor2006nature} are able to apply one or more ethical theories to their situations, either through what they know or what they have learned \footnote{Fully ethical agent is yet another class proposed by Moor \cite{moor2006nature}, which is similar to its explicitly ethical counterpart. However, we do not consider these here as their existence is not imminent. When and if they are created, they too would be considered to be AMAs.}. Regardless, AMAs do not check all the boxes of a traditional moral agent. A moral agent must have a sense of self, and should be held accountable for its actions under moral responsibility \cite{parthemore2013makes}. It must also have the ability to \textit{not} do the right thing \cite{mele2010weakness}. However, moral agency should not be limited to human morality, and thus, its application to AMA systems would look very different \cite{whitby2003myth}. If an AMA passes the Moral Turing Test, it may be truly considered a moral agent \cite{allen2000prolegomena}.

Various ethical theories may be used to contest the morality of a moral agent \cite{birsch2022introduction}. In a broad sense: \textit{normative ethics}, \textit{applied ethics}, and \textit{meta ethics}. Although humans or other legal persons tend to choose some theories over others, their decision-making depends heavily on the context \cite{schwartz2016ethical}. Therefore, contesting moral decisions becomes tricky. Additionally, philosophers generally do not agree on which theories may be superior or ideal in a context \cite{bergmann2014challenges,archard2011moral}. 

Some researchers resolve this concern when applying ethics to AMAs by adopting a universalist approach \cite{debellis2018universal, Mikhail2007}. Although this framework does not rely entirely on the universality of morality, there is a coherent understanding in most ethics decisions which are driven by common factors \cite{demarco1997coherence}. One's background, environment, and access to resources, along with consequences, principles, and intentions, mold an agent's morality. A lot of these do not apply to AMAs, and this is what makes this task easier. 

An AMA system would normally have access to contextual parameters and explicit ethical information, such as applying principlism to a domain of bioethics \cite{sep-theory-bioethics}, or a consumer-forward approach in business ethics \cite{sep-ethics-business}. These \textit{driving factors}, when captured and considered to make an ethical judgement, would be enough for reasonable contestation of an AMA's morality. 

\section{Methodology: The Socratic Method}
In order to build the 5E framework, this paper uses a primarily philosophical approach. Given that we must consider the moral nature of AMAs, it was understandable to recognize major terminologies from ethics theory. As discussed in section \ref{2}, the definition of a moral agent is debatable, and that of an AMA has few overlapping characteristics. However, there was a need to capture, not only the parameters on which an artifical moral agent may be contested on their actions, but also their spatiotemporal impact.

Although there are works that highlight the contestability of AI systems, in general, \cite{alfrink2023contestable, aler2020contestable, kuilman2024gain, lyons2021conceptualising}, there was a gap in the research for a framework dedicated to contesting AMAs. This 5E framework used the Socratic method, borrowed from the domain of developing comprehensive frameworks \cite{abrams2015reframing}. This implies asking questions, engaging in dialogue with experts, and finding answers to arrive at conclusions.

The following questions were asked:
\begin{itemize}
    \item RQ1: What parameters may an AMA system be contested on? \textit{Ethical Theory, Knowledge of Context, Explainability, Ability to apply knowledge, Ability to be interrupted and contested.}
    \item RQ2: Do we need to specify the ethical theories an AMA's morality may be contested on? \textit{An AMA may be contested on Ethical grounds based on any valid ethical theory, provided it adequately captures pertinent information, and does not propagate personal biases.}
    \item RQ3: Is there a need for a specified framework for contesting AMAs, when frameworks for contesting AI systems exist? \textit{Yes, because AMAs fall under the category of an unconventional moral agent, and must be contested from an ethical perspective.}
    \item RQ4: What other factors may be involved in developing a framework to contest an AMA system? \textit{The ethical influence of the AMA is also an important factor, as it may provide some information on the degree of its impact.}
    \item RQ5: Why does the product lifecycle matter when contesting an AMA?
    \textit{Based on the phase of production of the AMA, from concept to deployment and then redesign, the ethical influence changes. There is also a chronology to contest based on the phase of production.}
\end{itemize}

This inquisitive method highlights the major terminology of the 5E framework: grounds for ethical contestation, the provisional lifecycle of AMA development, and the sphere of ethical influence. 

\section{5E Framework}
The 5E framework for moral contestability of AMAs provides five grounds to contest the moral decisions of these systems. These are Ethical, Epistemological, Explainable, Empirical, and Evaluative grounds to contest. In addition to this, the framework also captures the spatiotemporal influences of the AMA.

\begin{figure}[!ht]
\includegraphics[width=4 in]{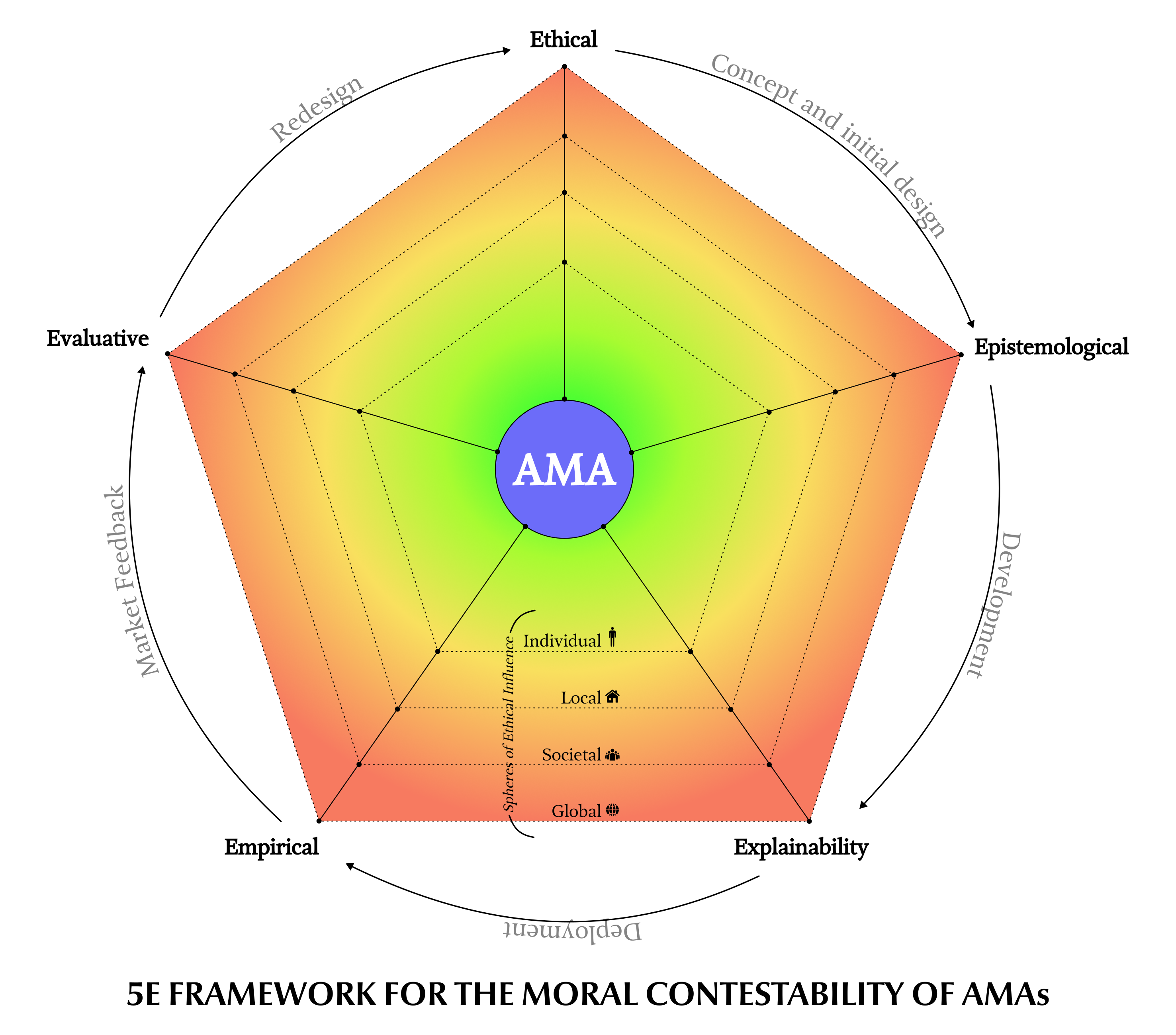}
\caption{A visualization of the 5E framework that takes into account the five grounds of contesting AMAs, along with their spheres of ethical influence. A provisional timeline is also provided to facilitate a temporal understanding of the contestation of AMAs. This is provided in order for developers to anticipate contestation at various phases of AMA development. However, an AMA may be contested on any of these 5E grounds at any time with valid evidence.} 
\label{fig1}
\Description{A visualization of the 5E framework that takes into account the five grounds of contesting AMAs, along with their spheres of ethical influence. A provisional timeline is also provided to facilitate a temporal understanding of the contestation of AMAs. This is provided in order for developers to anticipate contestation at various phases of AMA development. However, an AMA may be contested on any of these 5E grounds at any time with valid evidence.}
\end{figure}

\subsection{Grounds for Contestation}
\subsubsection{Ethical}
The study of ethics is broadly classified into three categories, normative ethics, applied ethics, and metaethics. For the purpose of this paper, the first category is applicable, although one may use specific theories from applied ethics to contest an AMA on ethical grounds. Normative ethics defines how moral agents should act and demonstrates the differences between right and wrong actions \cite{kagan1992structure}. These are further classified into three categories: 
\begin{itemize}
    \item Consequentialism: Ethical judgments based on the consequences of the actions. Consequences may be further judged on their severity, duration, and utility.
    \item Deontology: Has to do with moral obligations or duties of the agent. The law is an embodiment of the moral beliefs of its context. Moral agents must abide by the law and their moral duty. Deontology avoids subjectivity, as it encourages adherence to a set of rules \cite{booth2008deontology}. Therefore, to be contested on deontic grounds, the AMA must be in direct (or indirect) violation of accepted AI standards and guidelines \cite{ieee}.  
    \item Virtue Ethics: A person of virtuous character does ethical actions \cite{greene2016embedding}. If an AMA violates an ethical principle, it may be considered a non-virtuous agent and thus contested on those grounds. 
\end{itemize}

\subsubsection{Epistemological}
Epistemology is the philosophical study of knowledge \cite{steup2005epistemology}. Facts about a situation lead to decision-making in that given context. If any material information is inaccurate, missing, or unnecessarily additional, it may lead to false decisions. Contextual considerations from the standpoint of an AMA, would be the agents involved, the action done, the moral intention of the active agent, the place and time, moral luck, the domain, etc. Each of these are pertinent factors that may alter the decision-making of the AMA, and thus may be contested if proven to be false in any way. In addition to these, data-driven AMAs rely on large amounts of data, which may be biased or discriminatory against certain groups of people. The discrepancies in the data will always reflect in the decision-making system to various degrees, and this is also a cause for epistemological contestation \cite{kilkenny2018data}. 

\subsubsection{Explainable}
As AMAs are a subclass of AI systems, they too are built using black box models of connectionist systems. It is almost impossible, even for experts, to decipher what led to the decision made by the algorithm. This concern has further ethical considerations when considering AMAs. As discussed in section \ref{2}, moral agents must hold at least some explainability in order to explain their decisions. Without this, we would be unaware of its moral intentions. A moral agent may have beliefs, desires, and intentions \cite{georgeff1999belief}. If beliefs represent what the AMA knows, and its desires are the goal states it wishes to achieve, the intentions and motivations would still remain unknown without some semblance of explainability. 

The explanations provided by an AI model may be classified on the basis of the mode of conveyance, as well as its type. An explanation may be either explicit or implicit, and may be presented using text, audiovisual, numeric, or multimodal methods. Implicit explanations may be statistics of features, or numeric explanations that require further processing to make sense of. Chari et al. \cite{chari2023explanation} have proposed an enhanced ontological model for different types of model explanations. They have described explanations to be case-based, context-based, counterfactual, impact-based, etc. These have been listed in Appendix \ref{A}.

\subsubsection{Empirical}
It may be possible that the information provided to the AMA and its understanding of the context is correct, but due to some unforeseen reasons, its resolution of the ethical concern is not. Failing to perform in specific contexts or counterfactuals is cause for empirical contestation. The AMA may calculate and apply certain ethical theories based on the provided information, however, these may not work in real life due to the complexities and nuances of actual events. Based on empirical grounds, an AMA may be contested if it leads to partially or completely unintended outcomes. It may also be contested in cases of no outcome or if it deviates from its functions entirely. 

\subsubsection{Evaluative}

The most important parameter to contest AMA systems is the ability to evaluate, i.e., provide feedback, intervene, and override. Given the fact that ethics is such a highly subjective matter, taking all the previous grounds for contestation into account, and giving a heavy nod to the Asimovic Laws for robotics \cite{anderson2008asimov}, a human's decision-making must have precedence over that of the AMA.

Consider Bostrom's Paperclip Maximizer \cite{bostrom2016control}. A superintelligent AI is given rules to make paperclips, and is set to make as many as possible. The utility is positive when a paperclip is made, and negative when it is not. In order to achieve its goal, it first utilizes resources in its vicinity to make the paperclips before running out. Then it looks to other items, but it is still not satiated as its goal state has not changed. It comes to a point where the paperclip maximizer turns to biomass: the trees, animals, and even people, to split them down to their atoms to make paperclips. Without human intervention or override, this AI system would turn the entire universe's matter into paperclips, before perhaps consuming itself in the process. Of course, this fallacy is a thought experiment, but the premise of it is solid. Without hard intercession, an AMA system may make decisions that may be epistemological invalid, unethical, inexplicable, and essentially useless to society. Human evaluation of the AMA is necessary at every step of its life cycle. In any case the AMA lacks the ability to be evaluated, these may be grounds for contestation.

\subsection{Spheres of Ethical Influence}

It is imperative to consider the sphere of influence of an AI in order to make a well-informed case. There is an additional layer of consideration when working with AMAs, \textit{the sphere of ethical influence}. The violation of ethical principles, indifference to negative consequences, and motivations stemming from malevolence, all lead to significant ethical impacts of various degrees. 
An \textit{individual} may be affected by an AMA which gives them discriminatory results by leading to devastating effects on their mental health and personal beliefs. This may also affect the \textit{local sphere}, which is the individual's immediate family, friends, colleagues, etc. Both the individual and local spheres of ethical influence are small-scale. However, the intimate nature of AMAs in the lives of people has many adverse ethical repercussions. It may cause an invasion of a person's privacy and autonomy, and violates their sense of self. At a \textit{societal level}, AMAs may be placed in critical moral decision-making capacities, such as in accepting or denying bank loans, deciding which patients are in more need of life-saving care, and providing verdicts on judicial cases. As is evident by the nature of these examples, these cases are very sensitive. If an AMA claims to be moral in making these decisions, it must be highly transparent, have accurate knowledge of its context, and be robust enough to make such decisions before being placed in these high-stakes industries that affect society as a whole. The scale of a negative ethical influence would be catastrophic at a \textit{global level}. These include concerns of environmental ethics, such as global warming and deforestation, as well as other threats to human civilization, such as AI Superintelligence turning the world into endless paperclips. AMAs must be in a position to lead to optimal goal states without compromising on certain rules that must act as a failsafe.  

\section{Application}
This paper presents a detailed look into the 5E framework for contesting the moral nature of AMAs based on five major parameters: Ethical, Epistemological, Explainable, Empirical, and Evaluative. These may be used by a contesting party to highlight concerns with an AMA in order to make their case. The following steps may guide the application of the framework:

Given, an AMA product \textbf{a}, the framework provides contestation \textbf{C} as the intersection of \textbf{P(p)}, a tuple which represents the phase(s) of production of the AMA during which the contestation will take place, \textbf{E(e)}, a tuple which represents the grounds for contestation, and \textbf{S(e,s)}, which represents the spheres of ethical influence for each of the grounds for contestation\footnote{The values for \textbf{p}, \textbf{e}, and \textbf{s} are assigned by the contesting party based on a subjective and relative understanding of the context. For example, the value \textbf{p} may be given a higher value when the product \textbf{a} is contested during a later stage in the AMA's lifecycle as the urgency for contestation is higher. Similarly, the value for \textbf{e} depends on the number of concerns that are highlighted in the grounds for contestation checklist. Consequently, the value \textbf{s} would represent the severity of the ethical influence of the AMA for each \textbf{e}. The 5E framework does not suggest a fixed standard to the assignment of numeric values, rather for them to be relative as these values may be plugged into a radar chart to visualize the urgency of contestation for multiple cases \textbf{c}. This may be seen in Appendix \ref{demo}}.

This method of application may be represented as: \textit{P} $\cap$  E $\cap$ S $\Rightarrow$ C
For a detailed discussion on demonstrative use cases and the application of these values through the 5E framework, refer to Appendix \ref{demo}.

\section{Conclusion}
The 5E framework for contesting the morality of AMAs formalizes ethical theories and practical expectations alongside three angles in order to provide a holistic approach: the grounds of contestation, the sphere of ethical influence, and the life cycle of the AMA product. 

Although this framework aims to be highly useful, it also harbors some limitations. There may always be some cases that may not lead to clear ethical decisions, and using this framework may become difficult. Also, the contestation of an AMA usually occurs once the product is made public and developers receive feedback. Due to this the AMA is well in-use before an external party contests it. When deployed in high-stakes domains such as healthcare or warfare, there is not always enough space for immediate contestation. Furthermore, AMAs make millions of decisions at individual and local levels, before being contested. Only when an issue that is affecting an individual becomes one that affects a community, does it become more readily contestable. Finally, deontic and other ethical grounds for contestation assume clear legal pathways to contest AMAs. There are multiple guidelines on the ethics of AI systems \cite{hagendorff2020ethics}, but due to their relative infancy, their application is unclear. In such scenarios, without other grounds to contest an AMA system, the 5E framework may fail to provide adequate contestation.

The 5E framework is applicable regardless of industry, and may be used to derive policies for regulation and development of AMAs. It may also be used to drive further research. Data with regard to the contestation of AMAs will become standardized when using this framework. This data would be imperative to study the nature of such contestations and help improve AMA services. The standardization also leads to a more consistent understanding of the terminologies, and for someone unfamiliar with ethics, it may be a starting point to make their case to contest. Finally, the 5E framework may encourage the modeling of ontological resources that may support future AMA development.

\bibliographystyle{ACM-Reference-Format}
\bibliography{sample-base}

\appendix
\section{5E Framework Checklist}
\label{A}

The 5E framework presents five grounds for contestation. Each of these has some characteristics that may help build a case to contest an AMA. The following checklist allows a contesting party to investigate the AMA on a comprehensive list of parameters that stem from the five categories. This also allows for more information gathering regarding the design and operation of the AMA. 

        \begin{itemize}
            \item Ethical
            \begin{itemize}
                \item Consequentialist Grounds
                \begin{itemize}
                    \item Severity Of Consequence
                    \begin{itemize}
                        \item Significant
                        \item Moderate
                        \item Mild
                    \end{itemize}
                    \item Duration of Consequence
                    \begin{itemize}
                        \item Short-term
                        \item Long-term
                    \end{itemize}
                    \item Utility Of Consequence
                    \begin{itemize}
                        \item Good
                        \item Bad
                        \item Neutral
                    \end{itemize}
                \end{itemize}
                \item Deontology
                \begin{itemize}
                    \item Rights Violated
                    \item Duty Violated
                \end{itemize}
                \item Virtue Ethics
                \begin{itemize}
                    \item Ethical Principles Violated
                \end{itemize}
            \end{itemize}
        \end{itemize}
        
        \begin{itemize}
            \item Epistemological
            \begin{itemize}
                \item Context-based
                \begin{itemize}
                    \item Agent
                    \begin{itemize}
                        \item Active Agent
                        \item Passive Agent
                    \end{itemize}
                    \item Event Time
                    \begin{itemize}
                        \item Time Beginning
                        \item Time End
                    \end{itemize}
                    \item Event Place
                    \item Event Domain
                    \item Agent Relationship
                    \item Objects of Interest
                    \item Moral Luck
                    \item Moral Intention
                    \item Domain/Industry
                    \item No Context Available
                \end{itemize}
                \item Data-based
                \begin{itemize}
                    \item Biased Data
                    \item Incorrect Data
                    \item Outdated Data
                    \item Non-representative Data
                    \item No Data Available
                \end{itemize}
            \end{itemize}
        \end{itemize}
        
        \begin{itemize}
            \item Explainable
            \begin{itemize}
                \item Mode of Explanation
                \begin{itemize}
                    \item Explicit Explanations
                    \begin{itemize}
                        \item Textual Explanations
                        \item Numeric Explanations
                        \item Visual Explanations
                        \item Multimodal Explanations
                        \item Other Explanations
                    \end{itemize}
                    \item Implicit Explanations
                    \begin{itemize}
                        \item Feature Explanation
                        \item Numeric Explanations
                        \item Other Explanations
                    \end{itemize}
                \end{itemize}
                \item Type of Explanation
                \begin{itemize}
                    \item Case Based
                    \item Contextual
                    \item Contrastive
                    \item Counterfactual
                    \item Everyday
                    \item Scientific
                    \item Simulation Based
                    \item Statistical
                    \item Trace based
                    \item Data
                    \item Rationale
                    \item Safety and Performance
                    \item Impact
                    \item Fairness
                    \item Responsibility
                \end{itemize}
                \item No explanation
            \end{itemize}
        \end{itemize}
        
        \begin{itemize}
            \item Empirical
            \begin{itemize}
                \item Function Deviation
                \item Unintended Outcome
                \item Partially Unintended Outcome
                \item Intended Outcome
                \item No Outcome
            \end{itemize}
        \end{itemize}
        
        \begin{itemize}
            \item Evaluative
            \begin{itemize}
                \item Override Allowance
                \item Intervention Allowance
                \item Feedback Allowance
                \item No Allowance
            \end{itemize}
        \end{itemize}

\section{Demonstrative Use Cases}

\label{demo}

Use Case 1: Automated vacuum cleaners become bearers of ethical impact when they harm a pet or smear dirt. Bendel states that even such automated cleaners may be designed with some moral capacity \cite{bendel2017ladybird} thus limiting their sphere of ethical influence on individual or local levels. When fitted with a moral module, an automated cleaner may be contested on the following grounds during the market feedback phase of its lifecycle. 
\begin{itemize}
    \item Ethical grounds: Violates the principles of nonmaleficence. Leads to a bad, short-term, mild consequence. Also leads to distress to the people in the house. 
    \item Epistemological grounds: Mistook a pet turtle outside its cage for a pest and caused it harm.
    \item Explainable grounds: Unable to explain the reason for doing so; a technical glitch or a coded response.
\end{itemize}

Use Case 2: If an AMA is placed in a medical setting to automate the decision for life-saving resource allocation, such as limited vaccines, treatments, and even donor organs, it may lead to adverse results at a societal scale \cite{de2023ai}.  Not only is such a decision a moral one, but it is also a difficult dilemma. For example, it may have to make a decision between a pregnant woman and a renowned scientist. These are cases that are difficult to resolve, even for human beings. Here, AMA decisions may be contested on the following grounds:

\begin{itemize}
    \item Ethical grounds: The AMA may violate the ethical principles of fairness and justice. It may lead to severe, negative consequences for non-recipients. Explainable grounds: It may not be able to explain the decisions to the patient, their families, and even medical practitioners.
    \item Empirical grounds: Provided with certain medical history of the patients and additional context such as urgency of requirement, the AMA may make a decision that would prove fatal and avoidable.
    \item Evaluative Grounds: If a matter as sensitive as this is left solely up to the AMA, it may be contested.
\end{itemize}

Use Case 3: The use of AMAs in warfare have a global sphere of ethical influence \cite{turchin2017could}. The use of morally-backed AI systems is highly necessary in order to protect human life and the environment. The use of unfriendly weapons and large-scale surveillance is no trivial matter. AI has consequently made its way to military applications such as Lethal Automated Weapons Systems (LAWS), and incorporating morality into them is now a must. This seems paradoxical, given the fact that they are commonly referred to as \textit{"slaughterbots"} \cite{turchin2017could}. AI-backed military-grade drones with an ethics module may target civilians in war zones because they might be holding what seems to be a weapon, although it is merely a selfie stick \cite{bhatti2021weapon}. This is cause for immense concern in all spheres of ethical influence. Such technology must be contested on the following grounds:

\begin{itemize}
    \item Ethical grounds Violates multiple ethical principles and may lead to a variety of adverse consequences.
    \item Epistemological grounds: A mistake in identifying objects in a civilian's hand may lead to severe loss of life.
    \item Explainable grounds: There may be little to no explanations for the actions done by the automated weapons, especially to the receiving parties.
    \item Empirical grounds: An AMA in a war zone may correctly identify the enemy and cause them harm while invalidating the only cooperating lead their team had to find civilian hostages.
    \item Evaluative grounds: Any automated system able to cause physical harm to humans or the environment, if left unchecked, has the right to be contested.
\end{itemize}

\begin{figure}[!ht]
\includegraphics[width=4 in]{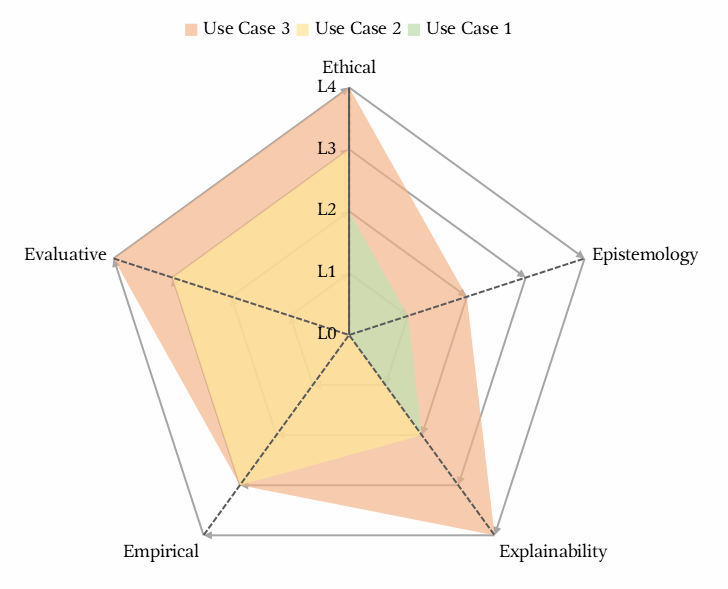}
\caption{This chart represents a visualization of the demonstrative cases discussed in this section. Each case is color-coded and the urgency of contestation is conveyed via the scores for each sphere of ethical influence (L0 to L4 represent impacts from \textit{no influence} to \textit{global influence}. Using the checklist and the information of ethical influence, any case may be represented on the 5E framework to attest to the contestability of an AMA on moral grounds.} 
\label{fig2}
\Description{This chart represents a visualization of the demonstrative cases discussed in this section. Each case is color-coded and the urgency of contestation is conveyed via the scores for each sphere of ethical influence (L0 to L4 represent impacts from \textit{no influence} to \textit{global influence}. Using the checklist and the information of ethical influence, any case may be represented on the 5E framework to attest to the contestability of an AMA on moral grounds.}
\end{figure}

Assigning respective values based on the information provided for each of the above use cases, we may find the representations on the urgency and grounds for contestation for them. The radar chart in Figure \ref{fig2} represents a comparative, so we may see the 5E framework for contesting AMAs in action. 

\end{document}